\documentclass[prl,aps,twocolumn,showpacs]{revtex4}
\usepackage{graphicx}
\usepackage{setspace} 
\DeclareGraphicsExtensions{.eps}

\def\E2w{\mbox{$E_{\mbox{{\tiny {2$\omega$}}}}$}}

\begin{document}

\title{Spectral signature of short attosecond pulse trains}

\author{E. Mansten}
\author{J. M. Dahlstr\"om}
\author{J. Mauritsson}
\author{T. Ruchon$^*$}
\author{A. L'Huillier}
\affiliation{Department of Physics, Lund University, P.
O. Box 118, SE-221 00 Lund, Sweden}
\author{J. Tate}
\author{M. B. Gaarde}
\affiliation{Department of Physics and Astronomy, Louisiana State University, Baton Rouge, LA 70803-4001}
\author{P. Eckle}
\author{A. Guandalini}
\author{M. Holler}
\author{F. Schapper}
\author{L. Gallmann}
\author{U. Keller}
\affiliation{Physics Department, ETH Zurich, Switzerland}

\begin{abstract}
We report experimental measurements of high-order harmonic spectra generated in Ar using a carrier-envelope-offset (CEO) stabilized 12\,fs, 800\,nm laser field and a fraction (less than $10 \%$) of its second harmonic. Additional spectral peaks are observed between the harmonic peaks, which are due to interferences between multiple pulses in the train. The position of these peaks varies with the CEO and their number is directly related to the number of pulses in the train. An analytical model, as well as numerical simulations, support our interpretation.

\end{abstract}


\maketitle

The availability of one, two or a few attosecond light pulses will enable physicists to pump and probe, and possibly to control coherently electronic processes at an unprecedented time scale. While attosecond pulse trains (APT), comprising many pulses, are easily produced by harmonic generation in gases~\cite{PaulScience2001,LopezMartensPRL2005}, the generation of single attosecond pulses (SAP), requires state-of-the-art laser systems and advanced characterization techniques, which to date, have been achieved by only two groups~\cite{KienbergerNature2004,SansoneScience2006}. Today, many laboratories in the world are working towards this goal. Different methods are being explored including generation with ultrashort laser pulses~\cite{KienbergerNature2004}, polarization gating~\cite{SansoneScience2006,SolaNP2006,MashikoPRL2008}, two-color frequency mixing~\cite{MashikoPRL2008,PfeiferPRL2006} and spatial filtering~\cite{GaardeOL2006,HaworthNP2007}. A common denominator to all of these methods is the requirement of stabilization of the laser CEO~\cite{TelleAPB1999, JonesScience2000, ApolonskiPRL2000}, which allows for the optimization of the electric field waveform required to generate one or two pulses~\cite{BaltuskaNature2003}.   

\begin{figure}[b]\centering
\includegraphics[width=\linewidth]{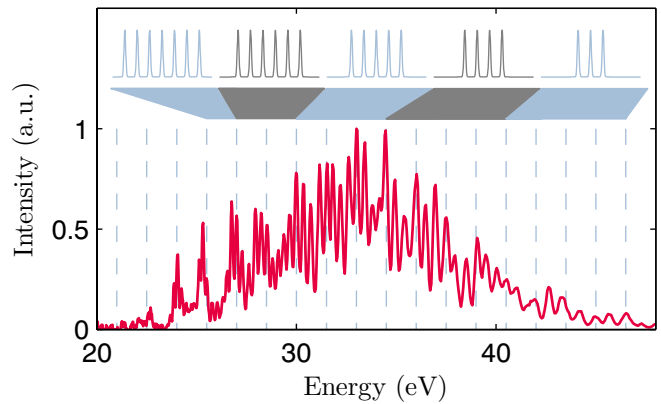}
\caption{(Color online) Experimental harmonic spectrum  The dashed light blue lines indicate the position of the harmonics, starting from the 14th. The top pictures illustrate the composition of the pulse train corresponding to a selection of a $\approx$3 eV spectral bandwith. The relative amplitude of the pulses is not known. \label{time_freq}}
\end{figure}

This Letter presents spectral measurements in conditions where the harmonic emission consists of only a few pulses. When the CEO is stabilized, we observe {\it additional frequency components} between the harmonics, the number of which decreases from low to high photon energies, as shown in Fig.~\ref{time_freq}. We provide a simple and straightforward interpretation
of the origin of these components, in terms of interferences between all of the emitted pulses in the train. We discuss how the
number of these extra peaks gives information about the number of pulses in the train depending on the frequency region considered. We argue that this is a general result, that will be met in all attempts to generate a few attosecond pulses
with phase-stabilized laser fields, and that can be used to get immediate information on the number of pulses present in the train as well as on the CEO-value needed for the production of a given number of pulses. The position of the additional frequency peaks varies with the CEO, in a way that depends on which trajectory (long or short) contributes to the harmonic emission~\cite{LewensteinPRA1995}.
  
Our work is closely related to two previous experimental results obtained in quite different conditions. Sansone and coworkers~\cite{SansonePRL2004} studied interferences between the long trajectories contributions of two consecutive harmonics and pointed out that these could also be interpreted as interferences between multiple pulses in the train. Pfeifer {\it et al.}~\cite{PfeiferOE2007} used ultrashort laser pulses and observed harmonic spectra containing harmonics and a continuous cut-off. By Fourier filtering they were able to remove the harmonic oscillation, thus revealing an additional modulation due to interferences between at least three pulses. In the present paper, multiple pulse interferences are observed {\it directly} and over the whole emitted spectrum, thus allowing us to get an immediate diagnostic on the status of the pulse train.

For our experimental studies we use a Ti:sapphire laser system delivering 30 fs CEO-stabilized 800 nm pulses at a repetition rate of 1 kHz with a maximum pulse energy of 0.8 mJ. The pulses are temporally compressed in a filament followed by chirped mirrors~\cite{HauriAPB2004}. The compressed 12\,fs laser pulses are sent through a 80\,$\mu$m type~I BBO crystal for second harmonic generation. 
The IR and the generated second harmonic radiation (blue) are separated in a dichroic interferometer where the relative phase, $\varphi$, between the IR and the blue can be finely controlled~\cite{MauritssonPRL2006}.
The two-color laser beam enters a vacuum chamber and is focused by a 50 cm radius of curvature spherical mirror into a pulsed argon jet. The Ar gas jet is positioned just after the laser focus for maximum efficiency as well as for phase-matching preferentially the short trajectory \cite{SalieresPRL1995}. The high harmonic emission, generated in the jet, enters an XUV-spectrometer composed of a platinum grating and a backside-illuminated CCD.  The acquisition of each harmonic spectrum is performed over a few thousand laser shots. 

Fig.~\ref{time_freq} presents the key result of this work: The harmonic spectrum generated by the few-cycle two-color driving pulse exhibits spectral peaks in addition to the even and odd harmonics. The number of extra peaks decreases as the harmonic frequency increases, while their amplitude relative to the harmonic peaks increases. Their origin can be understood through a {\it time domain} description of harmonic generation. Let us first consider for simplicity $n$ identical attosecond pulses, separated by a constant time interval ($T=2.7$ fs in the experiment), with the same complex amplitude [$a(t)$]. Their sum can be expressed as
\begin{equation}
s(t)=a(t) \otimes \sum^{j=n-1}_{j=0} {\delta(t-jT)},
\label{E1}
\end{equation}
and the Fourier transform of Eq.~(\ref{E1}) gives  
\begin{equation}
S(\Omega)=A(\Omega)  \sum^{j=n-1}_{j=0}  {e^{ij\Omega T}},
\label{E2}
\end{equation}
where $A(\Omega)$ is the spectral amplitude of each of the pulses which is equal to the Fourier transform of $a(t)$. The power spectrum is thus equal to
\begin{equation}
|S(\Omega)|^2=|A(\Omega)|^2 \left| \frac{\sin(n\Omega T/2)}{\sin(\Omega T/2)} \right| ^2
\label{E3},
\end{equation}
and consists of main peaks when the denominator in Eq.~(\ref{E3}) is zero, positioned at frequencies $\Omega=q\omega$, where $q$ is an integer and $\omega$ the laser carrier frequency. These are the harmonic peaks which, in the simple case where the periodicity is assumed to be a full laser cycle, can be odd or even. The spectrum in Eq.~(\ref{E3}) exhibits smaller peaks when the numerator is maximized, at the positions $(q+k/n+1/2n)\omega$ where $k$ is an integer between one and $n-2$ \cite{foot2}. The number of additional peaks between two harmonics is equal to $n-2$ and their intensity is of course reduced compared to the harmonics. Our model can be generalized to describe attosecond pulses with pulse-to-pulse varying amplitude and phase, formally expressed as 
\begin{equation}
S(\Omega)= \sum_{j} {A_j(\Omega) e^{ij\Omega T}}, 
\label{E4}
\end{equation}
where the number of pulses contributing is now included in the distribution of (complex) spectral components $A_j(\Omega)$. 
Based on this simple model, our interpretation of the results presented in
Fig.~\ref{time_freq} is the following: The spectral structure is the result of the coherent sum of a few attosecond pulses. By simply counting the number of fringes, we can determine the number of pulses in the train for a given energy range. As shown by the insets in Fig.~\ref{time_freq}, and as expected since the highest energy region requires the highest field amplitude, this number decreases with increasing energy. The importance of the additional frequency components relative to the main harmonic peaks reflects the intensity distribution of the pulses, their phase difference and the number of contributing pulses. 

The physics discussed here is quite general and by no means specific to a two-color field. Calculations performed with a one-color field and for realistic amplitude and phase variations between consecutive attosecond pulses also show additional frequency components for short enough APTs. The relationship between the number of these components and the number of pulses in the train is found to be remarkably robust. In our experiment, the addition of the blue field helps to observe the fringe structure simply because it reduces the number of pulses which interfere by half. 

\begin{figure}[b]\centering
\includegraphics[width=\linewidth]{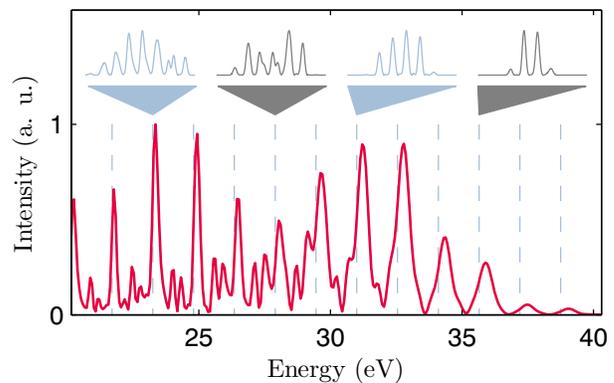}
\caption{(Color online) Calculated harmonic spectrum and structure of the pulse train at different energies. The position of the harmonics 14 to 25 is indicated by the dashed light blue lines. \label{mette}}
\end{figure}

To verify our interpretation, we have also performed numerical calculations based on the coupled solutions of the time-dependent Schr\"odinger equation and the Maxwell wave equation \cite{GaardeJPB2008}. Results shown in Fig.~\ref{mette} represent the contribution from the short trajectory only, which is selected by limiting the available return times in the single atom part of the calculation. The top of the figure indicates the structure of the pulse train at certain energies. The number of generated attosecond pulses increases towards lower energies and agrees reasonably well with the number of additional spectral components plus two. It confirms our interpretation that the frequency structure between the main harmonic peaks gives information on the time-frequency structure of the pulse train and more specifically on the number of pulses in the train contributing to a certain frequency range. This simple and direct method, which can be implemented ``on line", is complementary to more advanced characterization techniques like RABITT (Reconstruction of attosecond burst by interference of two-photon transition)~\cite{PaulScience2001} or FROG-CRAB (Frequency-resolved optical gating- Complete reconstruction of attosecond burst)~\cite{MairessePRA2005}.

\begin{figure}[t]\centering
\includegraphics[width=\linewidth]{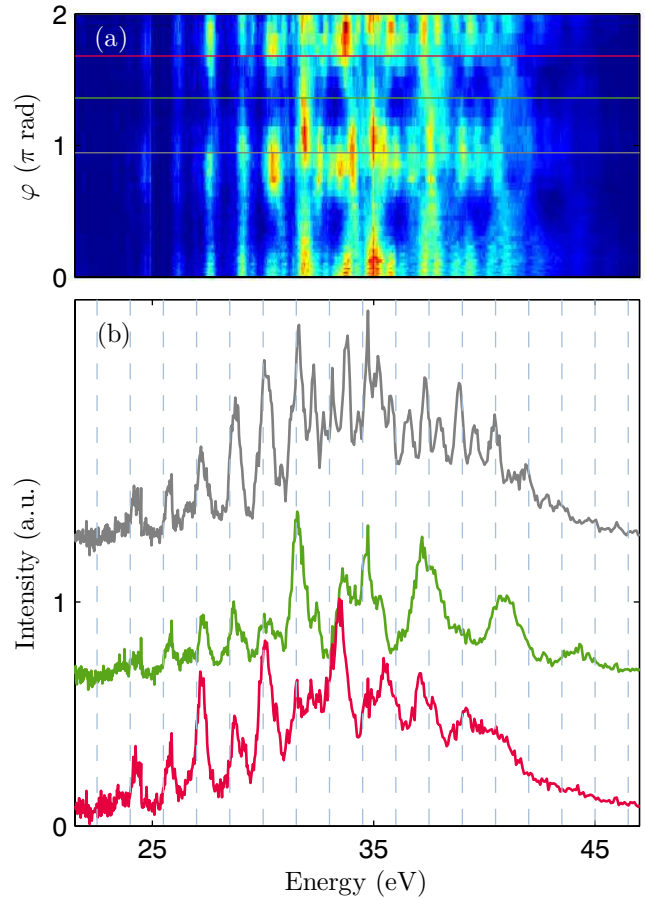}
\caption{(Color online) (a) Experimental harmonic spectrum as a function of the phase difference between the infrared and blue fields, $\varphi$. (b)~Spectra taken at the three phases marked in (a); the green and gray curves are shifted 0.6 and 1.2 intensity units respectively. The dashed light blue lines indicate approximately the positions of harmonics 15 to 31.\label{redblue}}
\end{figure}

The behavior of the measured harmonic spectra strongly depends on the relative IR-blue phase $\varphi$ as shown in Fig.~\ref{redblue}~(a). In general, the additional frequency components are best observed when $\varphi$ is such that even harmonics are as strong as odd harmonics [see gray curve in Fig.~\ref{redblue}~(b), corresponding to the delay outlined in gray in (a)], which leads in the time domain to a train of pulses with one pulse per cycle \cite{MauritssonPRL2006}. When two pulses per IR-cycle are generated, the harmonic signal is weaker and the high energy part of the experimental spectrum is dominated by the beating of two pulses separated by half an IR-cycle [see the green curve in Fig.~\ref{redblue}~(b)]. Finally, for the delay outlined by the red line, a continuum above 37 eV is observed. This is the spectrum of a short APT where one pulse is substantially stronger than the others and with higher energy components. Spectral filtering of this energy region (above 37 eV) is expected to yield a single attosecond pulse. The two-color scheme \cite{PfeiferPRL2006} proposed in the present experiment is of interest for single attosecond pulse generation since it does not require ultrashort laser pulses \cite{KienbergerNature2004} and since it should be more efficient than schemes based on polarization gating \cite{SansoneScience2006,SolaNP2006,MashikoPRL2008}.

\begin{figure}
\includegraphics[width=\linewidth]{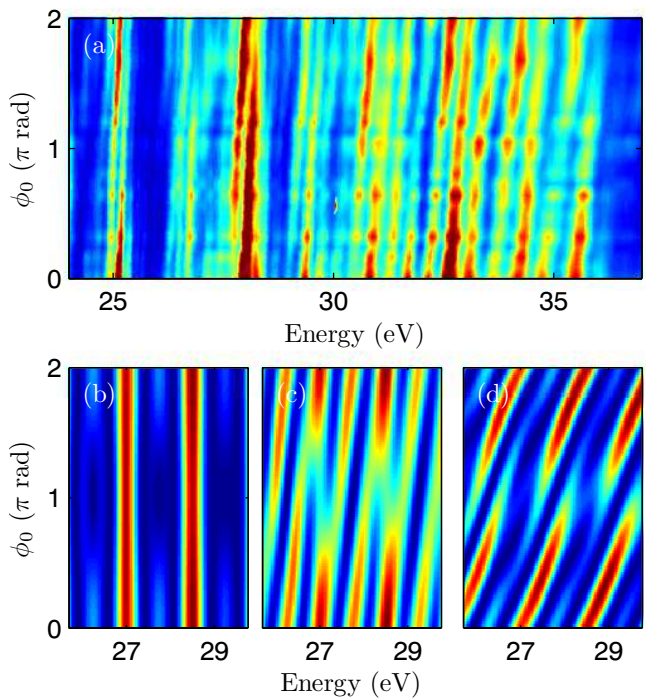}
\caption{\label{CEO}(Color online) (a) Experimental harmonic spectrum as a function of $\phi_0$. (b) Calculated harmonic spectra as a function of $\phi_0$ including amplitude effects. (c) Phase variations for the short trajectories added to the calculation in (b). (d) Phase effects corresponding to the long trajectories added to the calculation in (b).}
\end{figure}

In Fig.~\ref{CEO}~(a), we study the variation of the spectra as a function of the CEO phase ($\phi_0$). The position of the spectral structures changes with CEO, with a tilt increasing with energy. Indeed, when the CEO is not stable, the fringe structure is smeared out. Some of the fringes split or merge together as the CEO is changed. Finally, after a $2\pi$ change of $\phi_0$ the fringes come back to their original position. 
This periodicity can be understood with simple arguments. 
The spectral component $A_j(\Omega)$ corresponding to each attosecond pulse, which includes an amplitude and a phase reflecting the electron trajectory, depends on the value of the laser intensity ($I_j$) at which the attosecond pulse is created. When $\phi_0$ changes by $2\pi$, $I_j\rightarrow I_{j+1}$ and $A_j \rightarrow A_{j+1}$, so that $S(\Omega,\phi_0)=S(\Omega,\phi_0+2\pi)\exp{(-i\Omega T)}$. Consequently, the fringe pattern (proportional to $|S(\Omega,\phi_0)|^2$) varies periodically with $\phi_0$. 
The phase of $A_j(\Omega)$ can be approximated by $\alpha(\Omega) I_j$, where $\alpha$ depends on the spectral region as well as on the electron trajectory causing the emission \cite{LewensteinPRA1995}. The phase variation with intensity leads to a tilt of the interference fringes. For small phase variation, the position of the fringes can be shown to vary linearly with the CEO, with a slope proportional to $\alpha(\Omega)$ \cite{foot1}. The observed tilt of the bimaxima is thus a signature of the dipole phase variation during the laser pulse. How rapidly the fringes move depends on whether the harmonic emission is predominantly from the short or the long trajectory. A similar effect was observed by Sansone and coworkers and attributed, in their experimental conditions, to the long trajectory \cite{SansonePRL2004}. 
Finally, the variation of the amplitude $|A_j(\Omega)|$ affects the intensity distribution of the attosecond pulses, and even sometimes the number of pulses effectively contributing to the radiation, thereby changing the number of additional peaks. This effect can be seen experimentally as fringes splitting and merging as the number of pulses contributing changes with the CEO.  
To illustrate the different effects discussed above, we present in Fig.~\ref{CEO}~(b-d) the results of simple calculations based on Eq.~(\ref{E4}). Fig. \ref{CEO}~(b) shows the effect of the amplitude variation for a Gaussian 10~fs-long fundamental pulse, neglecting phase variation. The number of extra peaks oscillates between one and two (or one and zero at the highest energy) as a function of the CEO. The effect of both phase and amplitude variation is shown in (c,d) for $\alpha$ values corresponding to the short and long trajectories respectively, in the spectral region considered. Our experimental observation is consistent with the (rather slow) variation of the short trajectory.

In conclusion, we have studied high-order harmonic generation by short laser pulses, so that the attosecond pulse train comprises only a few pulses. In these conditions, the spectra do not include only harmonics of the laser frequency, but also extra peaks whose number decreases as a function of photon energy. These additional spectral peaks are due to multiple interferences between the pulses in the train. The number of additional peaks allows one to determine the structure of the attosecond pulse train as a function of energy. In addition the tilt of the fringes with carrier envelope phase gives us information on the dipole phase variation as a function of intensity, with a clear distinction between the short and the long trajectories.

\begin{acknowledgments}
This research was supported by the Marie Curie Early Stage Training Site (MAXLAS), the Integrated Initiative of Infrastructure LASERLAB-EUROPE within the 6th European Community Framework Programme, the Swedish foundation for international cooperation in research and high education (STINT), the Swedish Research Council, the National Science Foundation through grant no PHY-0449235, the Center for Computation and Technology at Louisiana State University, NCCR Quantum Photonics (NCCR QP) and a research instrument of the Swiss National Science Foundation (SNSF). 

\end{acknowledgments}
$^*$ Present address: CEA-Saclay, DSM, Service des Photons, Atomes et Molécules, 91191 Gif sur Yvette, France. 


\begin{thebibliography}{10}
\bibitem{foot2} The positions $k=0$ and $k=n-1$ do not lead to maxima, since they are not separated from the main peaks by a zero.
\bibitem{foot1} This can be formally shown by making a Taylor expansion of the fundamental laser intensity at the time where pulse $j$ is generated as a function of the CEO, and by inserting the expansion in Eq. (4).
\end{thebibliography}
\end{document}